\documentclass[letter,twocolumn]{jpsj3}
\usepackage{graphicx}
\usepackage{bm}
\usepackage{mathrsfs}
\usepackage{physics}

\usepackage{color}
\usepackage[normalem]{ulem}

\usepackage{comment}

\title{Microscopic Analysis of Lattice Distortion Effects in Rashba Systems}

\author{Yuuki Ogawa$^1$, Takumi Funato$^2$, and Hiroshi Kohno$^1$}
\inst{$^1$Department of Physics, Nagoya University, Nagoya 464-8602, Japan \\ 
$^2$Center for Spintronics Research Network, Keio University, Yokohama 223-8522, Japan \\ 
}	
\date{\today}
\abst{

 We theoretically study the effects of dynamical lattice distortion in a three-dimensional Rashba system 
based on a tight-binding model.  
 Considering an $sp$-electron system on a tetragonal lattice with broken inversion symmetry, 
 lattice distortions are incorporated through hopping-integral and crystal-axis modulations. 
 The effective Hamiltonian for the $p_z$- or $s$-derived band is found to have Rashba modulation terms, 
which induce various types of spin currents, such as Rashba, quadrupolar, perpendicular, 
and helicity currents. 
 The results are compared with those obtained by the method of local coordinate transformation.  
}

\date{\today}

\begin{document}

\maketitle

 Spin currents play central roles in spintronics applications, 
and to date, several generation methods have been established, 
including electrical\cite{SHEtheory, Sinova2015}, spin-dynamical\cite{Silsbee1979, Tserkovnyak2002}, 
and thermal\cite{Uchida2008} means. 
 Recently, a mechanical means was proposed based on the spin-vorticity coupling, which enables 
interconversion of angular momentum between mechanical rotation and electron spin.\cite{Matsuo2013} 
 Experiments have been done with shear flow in liquid metals\cite{Takahashi2016} 
and elastic rotational motion due to surface acoustic waves (SAW).\cite{Kobayashi2017} 
 Subsequently, it was pointed out that the spin-orbit interaction (SOI) can also contribute 
to the mechanical spin-current generation in heavy metals.\cite{Funato2018}
 A recent experiment using SAW suggests a spin-current generation through SOI,\cite{Kawada2021} 
but the results cannot be fully explained by these theories.
The SOI-mediated mechanism of mechanical spin-current generation still remains an open problem.

 For a spin-current generation using SAW, the Rashba SOI can be a primary SOI 
because of the lack of spatial inversion symmetry.\cite{Rashba1959}
 The Rashba SOI enables efficient spin-current generation because of the so-called spin-momentum locking. 
 The Rashba SOI is often considered as a relativistic effect of electrons moving in a net potential gradient. 
 On the basis of this picture, mechanical spin-current generation via Rashba SOI has been investigated.\cite{Funato2021H}
 Microscopically, however, the Rashba SOI emerges as an effective SOI 
in a band with multiorbital character with parity mixing.\cite{Nagano2009,Yanase2011} 
 This observation motivated us to examine the effects of lattice distortion in a Rashba system  
starting from a multiorbital tight-binding model.

\begin{figure}
        \centering
        \includegraphics[width=85mm]{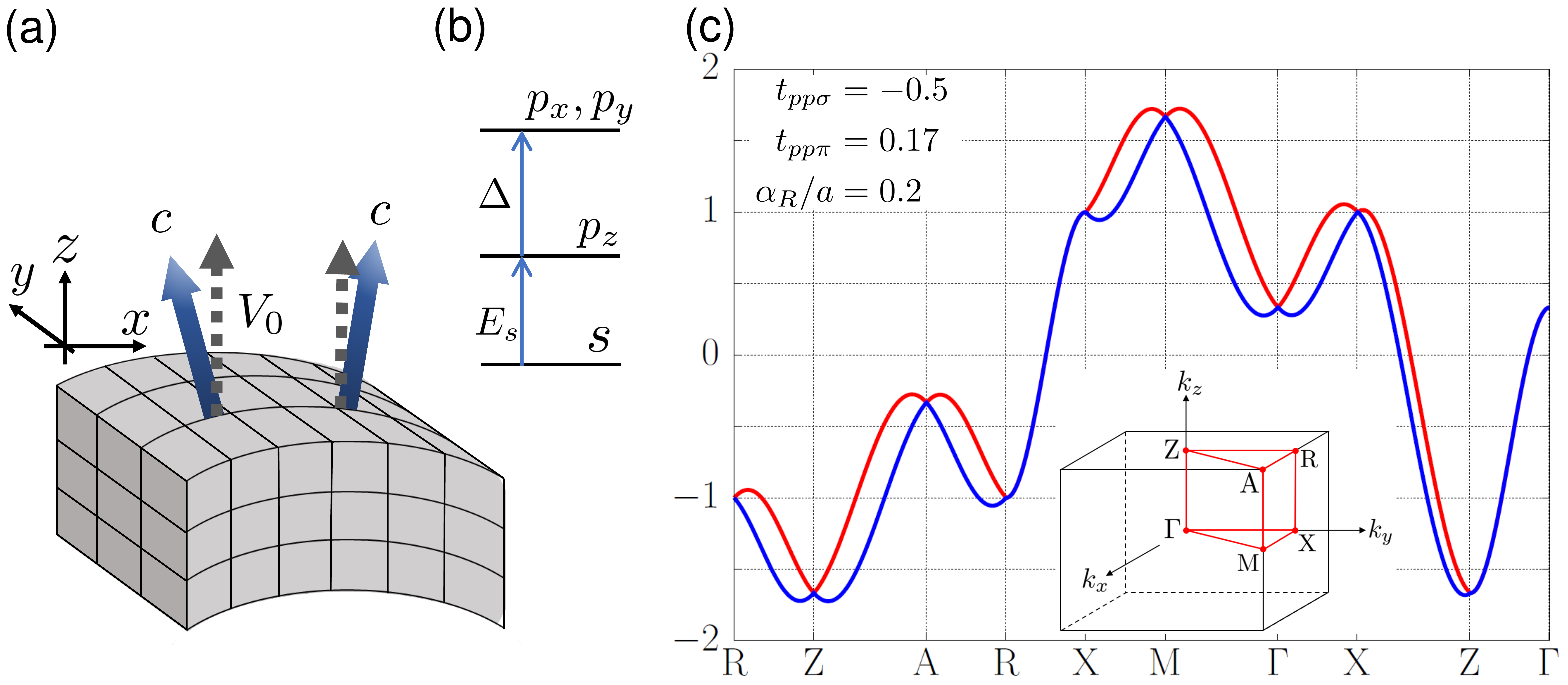}
        \caption{
 (a) Schematic illustration of dynamical lattice distortion induced, e.g., by SAW. 
 (b) Atomic level scheme of the starting tight-binding model. 
 (c) Energy dispersion of the $p_z$-derived band. 
} 
        \label{fig1}
\end{figure}

 In this Letter, we microscopically study the effects of lattice distortion in a three-dimensional (3D) 
Rashba system, and examine the spin-current generation. 
 We consider an $sp$-electron system on a tetragonal lattice with broken inversion symmetry.\cite{Yanase2011} 
 Lattice distortions are introduced 
via the modulation of hopping integrals and rotation of crystal axis (Fig.~\ref{fig1}(a)). 
 Assuming a level scheme as in Fig.~\ref{fig1} (b), we derive an effective Hamiltonian 
for the $p_z$- or $s$-derived band, 
which has a Rashba SOI (Fig.~\ref{fig1} (c)) and its modulations due to lattice distortion.  
 Classifying the interaction between lattice-distortion and electronic modes, 
we calculate spin currents generated by each of the coupling. 
 The model and the procedure we use closely follow an instructive review article,\cite{Yanase2011} 
in which a Rashba model is derived for an undistorted lattice.

We start with a tight-binding model of $s$ and $p$ electrons on a tetragonal lattice 
with broken inversion symmetry along the $c$ axis.\cite{Yanase2011} 
 The Hamiltonian is given by $H = H_{\rm kin} + H_{\rm CF} + H_{\rm odd} + H_{\rm so}$, with
\begin{align}
 H_{\rm kin} &= \displaystyle \sum_{<i,j>} 
 \sum_{m, n} 
   \left[  (E_{mn} + \delta E_{mn}) c_{im}^\dagger c_{jn} + {\rm H.c.}  \right]  ,   
\\
 H_{\rm CF}
 &= \displaystyle \sum_i  \left[ \Delta  (c_{ix}^\dagger c_{ix} +c_{iy}^\dagger c_{iy} ) 
    - E_s c_{is}^\dagger c_{is} \right]    , 
\\
 H_{\rm odd} &=V_0 \displaystyle \sum_i
( c_{is}^\dagger c_{ic} + {\rm H.c.} )  , 
\\
 H_{\rm so} &= \frac{\lambda}{2} \displaystyle \sum_i c_i^\dagger \, ({\bm l} \cdot {\bm \sigma} ) \, c_i  , 
\end{align}
which describe nearest-neighbor hopping ($H_{\rm kin}$), atomic levels including crystal-field splitting 
$\Delta >0$  ($H_{\rm CF}$), inversion symmetry breaking ($H_{\rm odd}$) with \lq\lq parity mixing'' $V_0$
(on-site mixing of $s$ and $p_z$ orbitals in undistorted lattice), and atomic spin-orbit coupling (SOC) ($H_{\rm so}$). 
 In $H_{\rm so}$, $\lambda$ is the atomic SOC constant, ${\bm l}$ is the orbital angular momentum, and 
$\vb*\sigma = (\sigma^x, \sigma^y, \sigma^z)$ is a vector of Pauli matrices.
 We defined a two-component operator, $c_{im} = {}^t (c_{im\uparrow}, c_{im\downarrow})$,  
that annihilates an electron in orbital $m$ ($=s,p_x,p_y,p_z$, or $s,x,y,z$ for short) at site $i$, 
and $c_i = {}^t (c_{is}, c_{iz} , c_{ix} , c_{iy} )$ is an eight-component operator. 
 The $p$ orbitals $m' = x,y,z$ are defined with reference to the laboratory frame, whereas 
$c_{ic}$ in $H_{\rm odd}$ is defined with reference to the crystal $c$ axis (Fig.~1 (a)).  
 The effects of lattice distortion are contained in $\delta E_{mn}$ and the tilting of the crystal $c$ axis. 
 We set $\hbar = 1$ throughout.

 Starting from this eight-component model, we successively eliminate 
\lq\lq high-energy'' bands to obtain a two-component Rashba model in a (dynamically) 
deformed lattice. 
 At each step, the Hamiltonian takes the form, 
\begin{align}
  H = \sum_{\bm k} c_{\bm k}^\dagger {\cal H} ({\bm k}) c_{\bm k} 
 + \sum_{{\bm k},{\bm q}} c_{{\bm k}+}^\dagger \delta {\cal H} ({\bm k}; {\bm q}) \, 
    c_{{\bm k}-} e^{-i\omega t}  ,
\end{align}
where $c_{\bm k}$ is the Fourier transform of $c_i$.  
 The first term is for the undistorted lattice, and the second describes the  
perturbation due to (dynamical) distortion. 
 The wave vector and frequency of the distortion are denoted by ${\bm q}$ and $\omega$, 
and we defined ${\bm k} \pm \equiv {\bm k} \pm \frac{\bm q}{2}$.

 The unperturbed $8 \times 8$ Hamiltonian is summarized as 
\begin{align}
\mathcal{H}(\vb*k)
&=\mqty(
\varepsilon_s - E_s  & V_0 + iV_z & iV_x & iV_y \\ 
 V_0 - iV_z  & \varepsilon_z & -i \lambda' \sigma^y  & i \lambda' \sigma^x \\ 
 -iV_x  & i \lambda' \sigma^y & \varepsilon_x + \Delta & -i \lambda' \sigma^z \\ 
 -iV_y & -i \lambda' \sigma^x & i \lambda' \sigma^z & \varepsilon_y  + \Delta \\ 
), \label{Eq.1}
\end{align}
with $\lambda' = \lambda /2$, $V_{m'} = - 2 t_{sp\sigma} \sin k_{m'} a$, and 
\begin{align}
\begin{cases}
\varepsilon_s = - 2t_{ss\sigma} (\cos k_x a  + \cos k_ya + \cos k_z a ) , 
\\
\varepsilon_z = - 2t_{pp\sigma} \cos k_z a - 2t_{pp\pi} ( \cos k_x a + \cos k_y a ) , 
\label{eq:energy}
\end{cases}
\end{align}
($\varepsilon_x$ and $\varepsilon_y$ are obtained from $\varepsilon_z$ by permutations), 
where $t_{mn\beta}$ are hopping integrals with $\beta$ ($= \sigma$ or $\pi$) symmetry 
between $m$- and $n$-orbitals at lattice spacing $a$. 
 For simplicity, we set $c = a$ for the lattice constants. 
This is a minimal microscopic model that produces an effective Rashba SOI  
starting from atomic SOC.\cite{Yanase2011}

 In a tight-binding model, lattice distortions are naturally introduced by local atomic displacements, $\delta \bm R$.  
 A nonuniform $\delta \bm R$ firstly modulates the hopping integrals in $H_{\rm kin}$, 
which we evaluate using the angular dependence given by Slater and Koster \cite{Slater1954} 
and assuming their inverse-square dependence on the interatomic distance (Harrison's rule).\cite{Harrison1980, Froyen1979} 
 To lowest order in $\delta \bm R$, the hopping integral between $m$- and $n$-orbitals, 
originally located at lattice points $\vb*r$ and $\vb*r +\vb*R$, respectively, is modulated by 
\begin{align}
\delta E_{mn}(\vb*r)
&= \vb*{\mathcal{E}}_{mn} (\vb*R) \cdot \frac{ \delta \vb*R (\vb*r +\vb*R) -\delta \vb*R (\vb*r)}{a} , 
\label{Eq.2} 
\end{align}
where $\vb*{\mathcal{E}}_{mn}$ are constant vectors that contain $t_{mn\beta}$ linearly;  
 for details, see Supplemental Material (SM).\cite{suppl}  
 We assume that the lattice distortions are of sufficiently long-wavelength, $q a \ll 1$ and $q \ll k_{\rm F}$, 
where $k_{\rm F}$ is the Fermi wave number, and adopt a continuum description, 
$\delta E_{mn} \simeq \mathcal{E}^i_{mn} \xi_{i,j} R_j / a$, 
where $\xi_{i,j} = \partial_j (\delta R^i)$ is the displacement gradient. 
 This leads to the modulation of $H_{\rm kin}$ with 
\begin{align}
\delta \mathcal{H}^{\rm (kin)}(\vb*k ; \vb*q)
&=\mqty(
\delta\varepsilon_s   & \delta V_z & \delta V_x  & \delta V_y  \\ 
 -\delta V_z & \delta\varepsilon_z  & \delta W_{zx} & \delta W_{yz}  \\  
 -\delta V_x & \delta W_{zx}  & \delta\varepsilon_x & \delta W_{xy}  \\  
 -\delta V_y & \delta W_{yz} & \delta W_{xy}  & \delta\varepsilon_y     
)  , 
\label{Eq.3}
\end{align}
where $(\delta\varepsilon_m , \delta V_{m'},  \delta W_{m'n'} ) 
= \sum_{\vb*R} ({\mathcal{E}}_{mm}^i , {\mathcal{E}}_{sm'}^i, {\mathcal{E}}_{m'n'}^i ) \, (R_j$ 
  $/a) (\xi_{i,j})_{{\bm q}, \omega} e^{i\vb*k\cdot\vb*R} $.\cite{suppl}

 In addition to hopping-integral modulations, lattice distortions may induce local rotation of crystal axes. 
 In our context, a local rotation tilts the axis of inversion symmetry breaking ($c$ axis) 
from the $z$ axis, and modulates $H_{\rm odd}$ 
through $c_{i c} \simeq  c_{i z}  + (\theta_i^y c_{i x} -  \theta_i^x  c_{i y})$, 
where ${\bm \theta}_i$ is the rotation angle  
$\vb*\theta(\vb*r ,t) =  \frac{1}{2} \grad \times \delta \vb*R(\vb*r ,t)$ 
at lattice site ${\bm r}_i$.
 This gives rise to a perturbation, 
\begin{align}
\delta \mathcal{H}^{\rm (rot)}(\vb*k ; \vb*q)
&= V_0 \mqty(
0 & 0 &  \theta_{\vb*q ,\omega}^y & - \theta_{\vb*q ,\omega}^x \\ 
0 & 0 & 0 & 0 \\ 
 \theta_{\vb*q ,\omega}^y & 0 & 0 & 0  \\  
- \theta_{\vb*q ,\omega}^x  & 0 & 0 & 0 ) . 
\label{Eq.5}
\end{align}
 (The suffices $\vb*q ,\omega$ denote Fourier components.) 
 The total perturbation is given by 
$\delta {\cal H} = \delta {\cal H}^{\rm (kin)} + \delta {\cal H}^{\rm (rot)}$.

\renewcommand{\arraystretch}{1.8}
\begin{table*}[t]
\begin{center}
\caption{Definition of lattice distortion modes, $\Xi_m$, and electronic modes, $\phi_m$, 
mutually coupled through $\delta \mathcal{H}_{\rm eff} $. 
 The induced spin currents are shown in the fifth row. 
 $\Xi_m$'s are expressed in terms of the strain tensor, $u_{ij} = \frac{1}{2} ( \xi_{j,i} + \xi_{i,j} )$,   
and the local rotation angle, $\theta^z = \frac{1}{2} (\xi_{y,x} - \xi_{x,y})$, around $z$ axis. 
 We introduced 
$({\bm k} \!\cdot\! \tau_x {\bm \sigma}) = (\tau_x)_{i\alpha} k_i \sigma^\alpha = k_x \sigma^y + k_y \sigma^x$, 
$({\bm k} \!\cdot\! \tau_z {\bm \sigma}) = (\tau_z)_{i\alpha} k_i \sigma^\alpha = k_x \sigma^x - k_y \sigma^y$, 
$({\bm k}_\parallel \!\!\cdot {\bm \sigma}) = k_x \sigma^x + k_y \sigma^y$, 
${\cal H}_\parallel^{(\beta)} ({\bm k}) = \frac{k_\parallel^2}{2m_\parallel} + \beta \alpha_{\rm R}  ({\bm k} \times {\bm \sigma})^z$, and  
$\frac{2}{\bar m} = \frac{1}{m_\parallel} + \frac{1}{m_z}$. 
 We also list $\phi_m'$ obtained by local coordinate transformation.\cite{Funato2021H,com1} 
 The bottom row shows the type of SAW (with propagation direction ${\bm q} \parallel \hat x$) 
that induces the respective spin currents (see Eqs.~(\ref{eq:js_Rayleigh}) and (\ref{eq:js_shear})). 
}
\begin{tabular}{|c||c|c|c|c|c|c|} \hline
 $m$ &  \  $1$  \  &  \  $2$  \  &  \  $3$  \  &  \  $4$  \  &  \  $(5, 6)$  \  &  \  $7$  \  
\\ \hline\hline 
\ $\Xi_m$ \ &  \ $u_{xx} + u_{yy}$ \ & \ $u_{zz}$ \  & \ $u_{xx} - u_{yy}$ \ & \ $u_{xy}$ \ & \ $(u_{xz}, u_{yz})$ \  &  \ $\theta^z$ \  
\\ \hline 
\ $\phi_m$ \ 
& $ {\cal H}_\parallel^{(1)} ({\bm k}) - \mu C_\parallel$ 
& $ \frac{{\tilde k_z}^2 }{m_z} - \mu C_z$ 
& $ \frac{k_x^2-k_y^2}{2m_\parallel} + \alpha_{\rm R}  ({\bm k} \!\cdot\! \tau_x {\bm \sigma} )$  
& $ \alpha_{\rm R} ({\bm k} \!\cdot\! \tau_z {\bm \sigma} )$
& $ - \alpha_{\rm R}  \tilde{k}_z (\hat z \times {\bm \sigma})$   
& $ \alpha_{\rm R} ({\bm k}_\parallel \!\!\cdot {\bm \sigma})$
\\ \hline 
\ $\phi_m'$ \ 
& $  {\cal H}_\parallel^{(1/2)} ({\bm k}) $ 
& $ \frac{ k_z^2 }{m_z} + \alpha_{\rm R} ({\bm k} \times {\sigma})^z $ 
& $ \frac{k_x^2-k_y^2}{2m_\parallel} + \frac{\alpha_{\rm R}}{2}  ({\bm k} \!\cdot\! \tau_x {\bm \sigma} )$  
& $ \frac{2 k_x k_y}{m_\parallel} - \alpha_{\rm R} ({\bm k} \!\cdot\! \tau_z {\bm \sigma} )$
& $ ( \frac{2k_z}{\bar m} - \alpha_{\rm R} \, \sigma^z \hat z \times ) \, {\bm k}_\parallel $  
& $ \alpha_{\rm R} ({\bm k}_\parallel \!\!\cdot {\bm \sigma})$ 
\\ \hline 
\ $ \langle j_{{\rm s},i}^\alpha \rangle $ \ 
& $ T_1 \, \epsilon_{i\alpha} $ 
& $ T_2 \, \epsilon_{i\alpha} $ 
& $ T_3 \, (\tau_x)_{i\alpha} $
& $ T_4 \, (\tau_z)_{i\alpha} $
& $ T_5 \, \delta_{iz} (-\delta_{\alpha y}, \delta_{\alpha x} ) $  
& $ T_7 \, \delta_{i\alpha}^\parallel $ 
\\ \hline 
\ SAW \ 
& Rayleigh  
& Rayleigh 
& Rayleigh  
& shear  
&  (Rayleigh, shear)   
& shear 
\\ \hline 
\end{tabular}
\label{table_I}
\end{center}
\end{table*}

 We assume a level scheme as shown in Fig.~1 (b), with $\Delta , E_s \gg  | t_{mn\beta}|$, 
and a chemical potential $\mu$ at around the $p_z$ or $s$ level. 
 We first eliminate high-energy $p_x$ and $p_y$ states by a canonical transformation, 
and then focus on the band of mainly $p_z$ (or $s$) character by a second transformation.
 The effective (unperturbed) Hamiltonian for the $p_z$ band thus obtained is 
\begin{align}
\mathcal{H}_{\rm eff} (\vb*k)
&=
\varepsilon_z  (\vb*k) + \frac{\alpha_{\rm R}}{a}  \Bigl[  \sin (k_x a)\sigma ^y -  \sin (k_y a)\sigma^x \Bigr]  
\label{eq:Rashba_C4v}
\\
&\simeq
E_{Z} + \frac{k _x^2 + k _y^2}{2m_{\parallel}} +\frac{\tilde k _z^2 }{2m_z}  + \alpha_{\rm R} (  k _x \sigma^y - k _y \sigma^x ) 
\label{eq:Rashba}
,
\end{align}
which has a Rashba term with  
$\alpha_{\rm R} \approx -2a V_0 t_{sp\sigma}\lambda/(E_s\Delta)$. 
 Note that $t_{sp\sigma} < 0$. 
 In the second line, we focused on the region near $\vb*Z = (0, 0, \pi/a)$, 
at which $\varepsilon_z (\vb*k)$ takes a minimum, $E_Z$ ($= -4 t_{pp\pi} + 2 t_{pp\sigma}$), 
and made an expansion with respect to $\tilde {\bm k}  \equiv \vb*k -\vb*Z$. 
 We defined $m_\parallel = (2t_{pp\pi} a^2 )^{-1}$ and 
$m_z = ( -2t_{pp\sigma} a^2 )^{-1} $.
 Note that $m_\parallel, m_z >0$. 
 The other band of mainly $s$ character has the same Rashba term but with opposite sign.\cite{Park2013} 
 In the following, we focus on the $p_z$ band at low electron density, 
as described by Eq.~(\ref{eq:Rashba}), but still with $k_{\rm F} \gg q$. 
 Also, the electron energy (including the chemical potential $\mu$) is measured from $E_Z$. 
 Note that Eq.~(\ref{eq:Rashba}) has a higher point-group symmetry (C$_{\infty v}$) 
than Eq.~(\ref{eq:Rashba_C4v}), which has only C$_{4v}$.

 The same procedure, but now including $\delta \mathcal{H}^{\rm (kin)}$ and $\delta \mathcal{H}^{\rm (rot)}$,  
leads to an effective $2\times 2$ perturbation,\cite{suppl}    
\begin{align}
  \delta \mathcal{H}_{\rm eff} (\vb*k; \vb*q) 
= - \sum_{m=1}^7 \phi_m ({\bm k}) \, \Xi_m ({\bm q}) \, e^{-i\omega t}   , 
\label{eq:dH_eff}
\end{align}
where $\Xi_m$ are lattice deformation modes and $\phi_m$ are electronic modes defined in Table I. 
 Some remarks are in order. 
 First, every term in $\delta {\cal H}_{\rm eff} $ can find its origin 
either in the kinetic energy or in the Rashba SOI in Eq.~(\ref{eq:Rashba}),  
and the latter may be called Rashba modulation terms. 
 Second, the undeformed Rashba model, Eq.~(\ref{eq:Rashba}), couples the spin to the in-plane wave vector 
${\bm k}_\parallel = (k_x, k_y, 0)$, but the strains $\Xi_5$ and $\Xi_6$ introduce the coupling to $\tilde k_z$.
 Third, there arises a term $\tilde k_z ({\bm \theta} \cdot {\bm \sigma})$ from each of  
$\delta {\cal H}^{\rm (kin)}$ and $\delta {\cal H}^{\rm (rot)}$, but they were canceled with each other. 

 Last but not least, $\phi_1$ and $\phi_2$ contain constant terms (${\bm k}$-independent unit matrices), 
$C_\parallel = \frac{2}{m_\parallel a^2 \mu}$ ($\equiv C_\parallel^0$) and $C_z = \frac{2}{m_z a^2 \mu}$ ($\equiv C_z^0$), 
arising from the expansion, such as $t_{pp\pi} u_{xx} \cos k_x a = t_{pp\pi} u_{xx} [ 1 - O((k_x a)^2)] $, 
thus describing band-width modulations. 
 For a low but fixed electron density, they should be absorbed into the (local) chemical potential, 
$\mu ({\bm r},t) = \mu + \delta \mu ({\bm r},t)$. 
 More appropriately, we impose the local charge neutrality 
through a chemical potential modulation, $\delta \mu$. 
 We find\cite{suppl} 
\begin{align}
 \frac{\delta \mu}{\mu} 
= \left( C_\parallel^0 - \frac{n - n_1}{\mu \nu} \right) (u_{xx} + u_{yy}) + \left( C_z^0 - \frac{n}{\mu\nu} \right) u_{zz}  , 
\end{align}
with electron density $n$, the (band-averaged) density of states $\nu = \nu (\mu)$, and 
$n_1 = \frac{ \mu m_\parallel \sqrt{2\mu m_z}}{(2\pi)^2} \chi ( 1 + 2 \chi^2) \arctan \chi$,  
where $\chi = \alpha_{\rm R}/v_{\rm F}$ with $v_{\rm F} = \sqrt{2\mu/m_\parallel}$ 
is a dimensionless Rashba constant.\cite{suppl} 
 Since such screening occurs locally and instantaneously compared to the scales of SAW,  
we include $\delta \mu$ in $\delta {\cal H}_{\rm eff}$ 
and treat them as a total perturbation. 
 This amounts to redefining $C_\parallel$ and $C_z$ as\cite{suppl,com2}    
\begin{align}
 C_\parallel &= \frac{n - n_1}{\mu \nu}  , \ \ \ \ \  
 C_z = \frac{n}{\mu \nu}  . 
\label{eq:C_perp,C_z}
\end{align}
 Note that the original $C_\parallel^0$ and $C_z^0$ have disappeared.

 Before proceeding, we compare the present result with those derived by the method of local 
coordinate transformation starting from the isotropic Rashba model, Eq.~(\ref{eq:Rashba}),\cite{Funato2021H,com1}  
\begin{align} 
 \delta {\cal H}' ({\bm k}; {\bm q})
&=  - \sum_{m=1}^9 \phi_m' ({\bm k}) \, \Xi_m ({\bm q}) \, e^{-i\omega t}  , 
\label{eq:dH_coord_tr}
\end{align}
where 
$\phi_m'$ are given in Table I, and we have added  
$(\Xi_8,\Xi_9) = (\theta^x,\theta^y)$, and 
$(\phi_8', \phi_9') = - [\, \alpha_{\rm R} \sigma^z + (\frac{1}{m_\parallel} - \frac{1}{m_z}) \, k_z \hat z \times ] \, {\bm k}_\parallel $.
 Among several differences from Eq.~(\ref{eq:dH_eff}), the followings may be notable.  
 First, while the combination $\phi_3' \Xi_3 + \phi_4' \Xi_4$ is invariant under C$_{\infty v}$, 
 $\phi_3 \Xi_3 + \phi_4 \Xi_4$ in Eq.~(\ref{eq:dH_eff}) is not; 
the latter is invariant only under C$_{4v}$,\cite{com3} reflecting the symmetry 
of the starting (tight-binding) model. 
 This means that our result, Eq.~(\ref{eq:dH_eff}), cannot be obtained if the isotropic Rashba model, 
Eq.~(\ref{eq:Rashba}), is taken as a starting model. 
 Second, the terms with $k_z$ arise from the kinetic energy ($\sim k_z k_j u_{zj}$ in $\phi_5', \phi_6'$), 
not from the Rashba coupling. 
 Finally, there are no constant terms initially in $\phi_1'$ and $\phi_2'$. 
 However, imposing the local charge neutrality will produce $C_\parallel' = C_z' = n/\mu \nu$.

\begin{figure}
\begin{center}
 \centering
   \includegraphics[width=9.0cm]{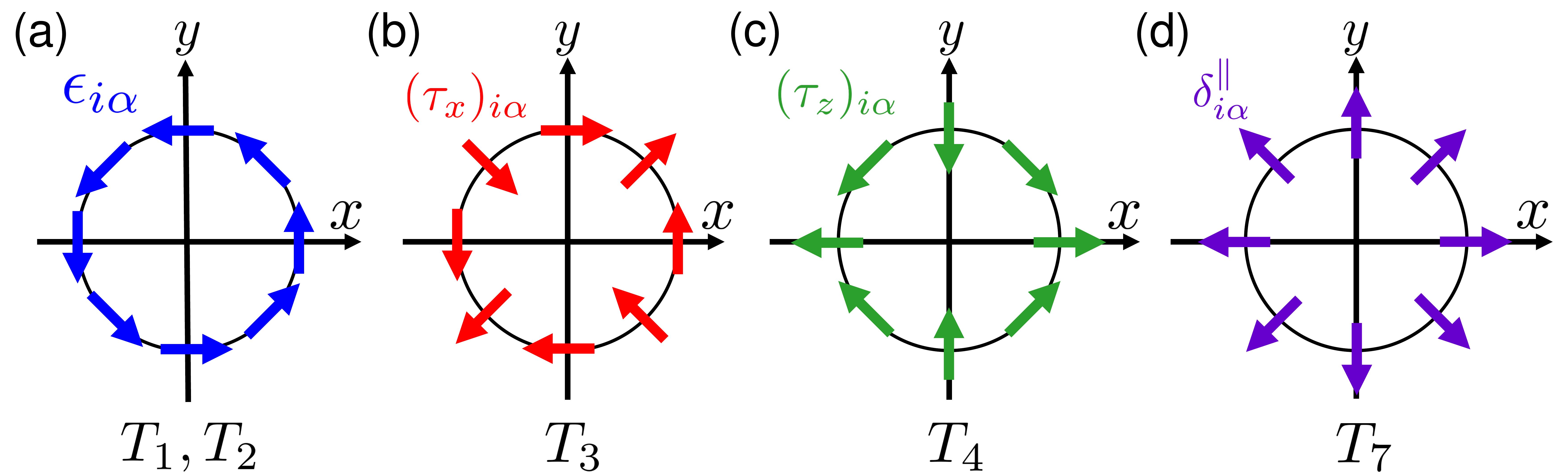}
\caption{2D spin-current patterns generated by lattice distortions. 
 Each arrow indicates spin direction, and its position viewed from the origin indicates flow direction. 
(a) Rashba spin current. (b,c) Quadrupolar spin currents. (d) Helicity current.}  
\label{spin_current}
\end{center}  
\end{figure}

\begin{figure}
\begin{center}
 \centering
   \includegraphics[width=7.5cm]{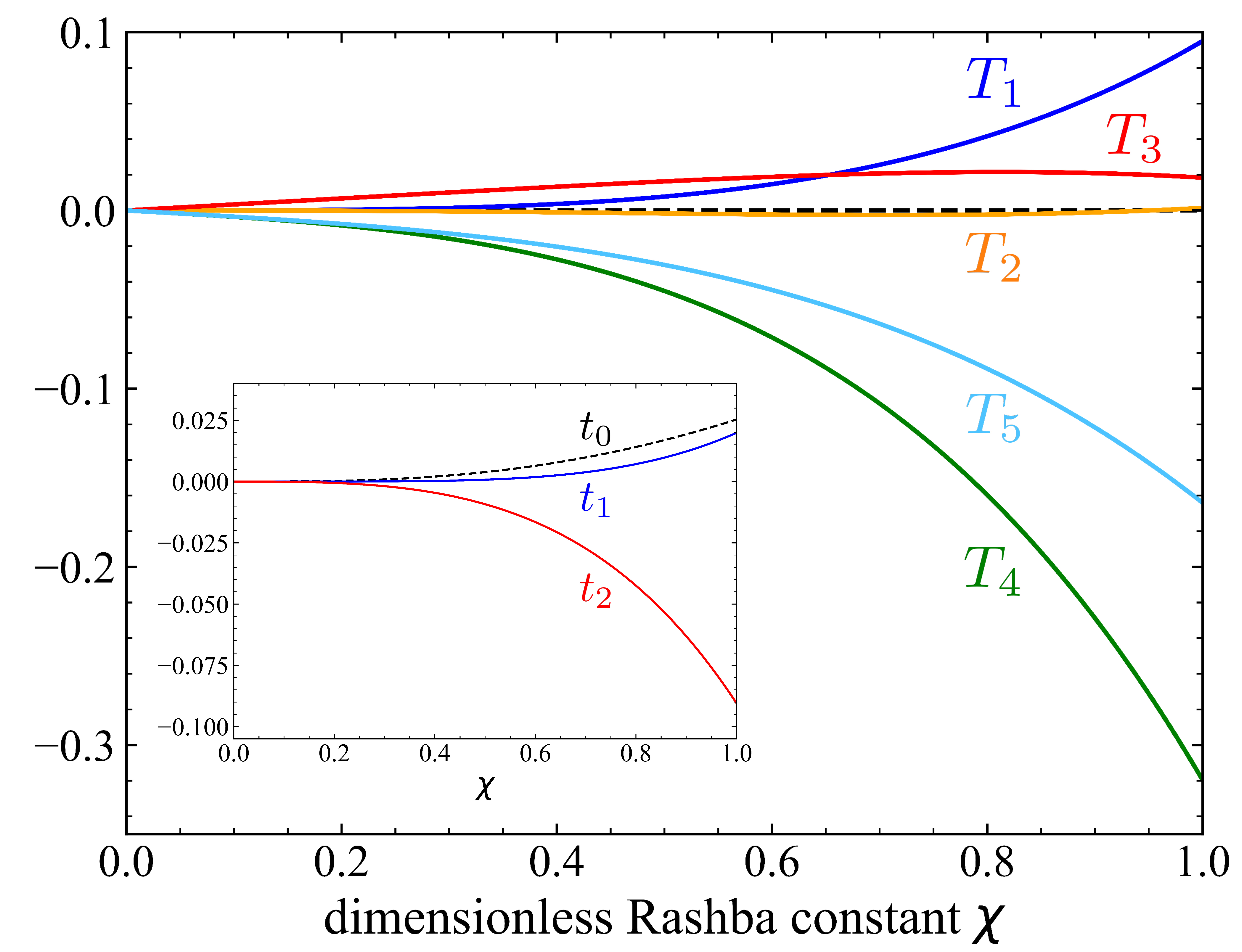}
\caption{Plots of $T_n$ as functions of 
$\chi = \alpha_{\rm R} m_{\parallel}  / \sqrt{2 m_{\parallel} \mu }$ for $\mu > 0$. 
 The inset shows $t_0$, $t_1$, and $t_2$, 
which constitute $T_1 = C_\parallel \, t_0 + t_1$ and $T_2 = C_z \, t_0 + t_2$. 
 The smallness of $T_2$ is due to near cancellation of the two terms 
 ($C_z \simeq 3.6$ at $\chi = 1$\cite{suppl}). 
  The unequal $T_3$ and $T_4$ reflect the C$_{4v}$ symmetry.} 
\label{graph}
\end{center}  
\end{figure}

 Let us now calculate (nonequilibrium) spin currents in response to $\delta {\cal H}_{\rm eff}$ 
using Kubo formula. 
 The spin-current operator is given by 
$\hat j_{s,i}^\alpha (\vb*q) = \sum_{\vb*k} c_{\vb*k -}^\dagger j_{s,i}^\alpha c_{\vb*k +}$ with
\begin{align}
 j_{s,i}^\alpha
\equiv\frac{1}{2} \left\{ \pdv{\mathcal{H}_{\rm eff}(\vb*k)}{k^i}  , \sigma^\alpha \right\}_+ 
=  \frac{\tilde k _i}{m_i} \sigma^\alpha + \alpha_{\rm R} \epsilon_{i\alpha}  ,  
\label{Eq.12}  
\end{align}
where $\alpha$ specifies the spin direction and $i$ the flow direction. 
 The velocity $\partial \delta \mathcal{H}_{\rm eff} / \partial k^i$ from the distortion part 
is dropped since it contributes only to equilibrium spin currents. 
 Focusing on the terms linear in $\omega$ (i.e., response to $\dot \Xi = \partial \, \Xi / \partial t$), 
we calculate the so-called Fermi-surface terms (other terms turned out to vanish), 
\begin{align}
 \langle \hat j_{s,i}^\alpha (\vb*q) \rangle_{\omega}
=  \frac{i\omega}{2\pi V} \sum_m \sum_{\vb*k} 
    \tr[ j_{s,i}^\alpha G^{\rm R}_{\vb*k} \phi_m G^{\rm A}_{\vb*k} ] \Xi_m ({\bm q}) , 
\label{Eq.13} 
\end{align}
where $G^{\rm R}_{\vb*k}$ ($G^{\rm A}_{\vb*k}$) 
is the retarded (advanced) Green's function (see SM\cite{suppl}). 
 We introduced a  $\delta$-function impurity potential and evaluated the scattering time $\tau$ 
in the Born approximation. 
 Assuming good conductivity (long $\tau$),\cite{com4} we retain the leading contributions. 
 The result is\cite{suppl}   
\begin{align}
  j_{{\rm s},i}^\alpha ({\bm r}, t) /  j_{s0} \tau 
&=  \epsilon_{i\alpha} \bigl[ T_1 (\dot{u}_{xx} + \dot{u}_{yy})  
   + T_2 \dot{u}_{zz} \bigr] 
\label{eq:js}
\\
& \ \ 
    + T_3  (\tau_x)_{i\alpha} (\dot{u}_{xx} - \dot{u}_{yy}) + T_4 (\tau_z)_{i\alpha}  \dot{u}_{xy}  
\nonumber \\
& \ \ 
   + T_5 \delta_{iz} (\delta_{\alpha y} \dot{u}_{xz}  -  \delta_{\alpha x} \dot{u}_{yz} ) 
    + T_7 \, \delta_{i\alpha}^\parallel \, \dot \theta^z  , 
\nonumber
\end{align}
where $j_{s0} = 2 \mu ^2 \sqrt{ m_\parallel m_z}$. 
 Together with $\epsilon_{i\alpha}$ and 
$\delta_{i\alpha}^\parallel = \delta_{ix}\delta_{\alpha x} + \delta_{iy}\delta_{\alpha y}$, 
we used the Pauli matrices $\tau_x$ and $\tau_z$ to express 
the spin and flow directions in the 2D $xy$ plane ($i, \alpha = x,y$), 
$(\tau_x)_{i\alpha}  = \delta_{ix}\delta_{\alpha y} + \delta_{iy}\delta_{\alpha x}$, 
and 
$(\tau_z)_{i\alpha}  = \delta_{ix}\delta_{\alpha x} - \delta_{iy}\delta_{\alpha y}$. 
 Such \lq\lq spin-current patterns'' are illustrated in Fig.~\ref{spin_current}. 
 The coefficients $T_n$ are dimensionless functions of $\chi$, 
which are given in SM \cite{suppl} and plotted in Fig.~\ref{graph}. 
 In real materials, $\chi$ can be of order unity.\cite{Ishizaka2011}

 In Eq.~(\ref{eq:js}), the $T_1$- and $T_2$-terms arise from (dynamical) distortions 
that preserve the symmetry of the undistorted lattice (i.e., $\dot{u}_{xx} + \dot{u}_{yy}$ or $\dot{u}_{zz}$). 
 They share the same spin-current pattern as the equilibrium spin current 
($\sim \epsilon_{i\alpha}$)\cite{Rashba2003}
in the undistorted Rashba system (Fig.~\ref{spin_current} (a)), and may be termed a \lq\lq Rashba spin current.'' 
 The $T_3$- and $T_4$-terms arise from the in-plane shear distortions 
($\dot{u}_{xx} - \dot{u}_{yy}$ or $\dot{u}_{xy}$). 
 They exhibit quadrupolar patterns (Fig.~\ref{spin_current} (b,c)), 
thus termed \lq\lq quadrupolar'' spin currents. 
 The $T_5$-terms, induced by vertical shear distortions (containing the $z$ axis), 
flow in $z$ direction with spin normal to the shear ($yz$ or $xz$) plane. 
 These may be termed \lq\lq perpendicular'' spin currents. 
 Curiously, $\dot \theta^z$ induces no spin currents, $T_7 = 0$, at $O(\tau)$ and $O(\tau^0)$.  
 However, in this process, the energy scale to be compared with $1/\tau$ is not the Fermi energy 
but $\alpha_{\rm R} k_{\rm F}$. 
 If we look at the region, $\alpha_{\rm R} k_{\rm F} \tau \ll 1$, in which the Rashba splitting 
is smeared by the broadening, we find $ T_7 = - \chi/(3\pi^2)$, 
and a \lq\lq helicity current'' (Fig.~\ref{spin_current} (d))\cite{com5} is induced at $O(\tau)$ 
by the lattice vorticity field $\dot \theta^z$.

 Let us apply the above result to two types of SAW. 
 For a Rayleigh wave applied in the $x$ direction (Fig.~\ref{fig1}(a)), 
$\delta {\bm R} = (\delta R_x, 0, \delta R_z) \, e^{i(qx-\omega t)} e^{q'z}$, we have  
\begin{align}
  j_{{\rm s},i}^\alpha 
&=  j_{s0} \tau 
\Bigl\{ \epsilon_{i\alpha}  [\, (T_1 + \eta_i T_3 ) \dot{u}_{xx} + T_2 \dot{u}_{zz} ] 
          +T_5 \delta_{iz}   \delta_{\alpha y} \dot{u}_{xz}\Bigr\} . 
\label{eq:js_Rayleigh}
\end{align}
 The factor $\eta_x = - \eta_y = 1$ shows that 
$T_3$ introduces in-plane anisotropy to the (isotropic) $T_1$-contribution.  
(This picture is valid for $|T_1| > |T_3|$.)  
 A shear wave applied in the same direction, 
$\delta {\bm R} \propto \hat y \, e^{i(qx-\omega t)} e^{q'z}$, yields 
\begin{align}
  j_{{\rm s},i}^\alpha 
&= \frac{1}{2}  j_{s0} \tau 
  \Bigl\{ (\eta_i T_4 + T_7) \, \delta_{i\alpha}^\parallel  \, \dot \xi_{y,x}  
             - T_5 \, \delta_{iz} \delta_{\alpha x} \dot \xi_{y,z}   \Bigr\} , 
\label{eq:js_shear}
\end{align}
which consists of the quadrupolar ($\propto T_4$), helicity ($\propto T_7$), 
and perpendicular ($\propto T_5$) currents. 
 The first two interfere and make the in-plane flow anisotropic. 
 The magnitude of these spin currents is proportional to $\omega^2$. 
 Interestingly, these two cases exhaust our spin-current patterns; see Table I, bottom row.

 The perpendicular spin current in Eq.~(\ref{eq:js_Rayleigh}), $j_{\rm s}^\perp $, 
has the same spin-current pattern as the one induced by the spin-vorticity coupling, 
$j_{\rm s}^{\rm svc} = (\nu \omega / 2q) \, \dot \theta^y$,\cite{Matsuo2013}  
thus it may be detected by a similar method.\cite{Kobayashi2017}  
 To estimate the magnitude, we may use the values of BiTeI ($\mu = 0.2$eV, $m = 0.09 m_e$, and $\chi = 0.71$) \cite{Ishizaka2011} 
for $j_{\rm s}^\perp $,
and $\nu \sim 2 \times 10^{22}$/eVcm$^3$ (Cu) and 
$\omega / q \sim 4$km/s (Rayleigh wave) for $j_{\rm s}^{\rm svc}$, and obtain 
$ (j_{\rm s}^\perp )_{\rm BiTeI} / (j_{\rm s}^{\rm svc})_{\rm Cu}  
  \simeq  \mu \tau (u_{xz}/\theta^y)$. 
 Thus, for $\mu \tau \gtrsim 1$, 
$j_{\rm s}^\perp $ can be comparable in magnitude with $j_{\rm s}^{\rm svc}$. 
 On the other hand, detecting the 2D spin currents (Fig.~\ref{spin_current}) 
may require some novel idea, 
but it will provide an unambiguous proof of the present (Rashba-induced) mechanism.

 In summary, we have studied the effects of lattice distortion in a 3D Rashba system 
starting from a multi-orbital tight-binding model with hopping-integral and crystal-axis modulations. 
 By eliminating high-energy bands, we obtained an effective Rashba Hamiltonian 
perturbed by strains and local rotation. 
 The perturbation Hamiltonian may be viewed to arise as modulations of the unperturbed part, 
but it does not necessarily inherit the symmetry of the latter. 
 We found a new Rashba modulation term due to vertical shear strain, 
which is proportional to $\tilde k_z$ and induces a spin current in $z$ direction, 
as well as others that induce various 2D spin currents.

\small

 This work is supported by JSPS KAKENHI Grant Numbers JP19K03744 and JP21H01799. 
 Y.~O. would like to thank the \lq\lq Nagoya University Interdisciplinary Frontier Fellowship'' supported by Nagoya University and JST, the establishment of university fellowships towards the creation of science technology innovation, Grant Number JPMJFS2120.


\normalsize


\begin{thebibliography}{99}
  \bibitem{SHEtheory} M. I. D'yakonov and V. I. Perel', Zh. Eksp. Teor. Fiz. Pis. Red. {\bf 13}, 657 (1971)
[Sov. Phys. JETP Lett. {\bf 13}, 467 (1971)].
  \bibitem{Sinova2015} J. Sinova, S. O. Valenzuela, J. Wunderlich, C. H. Back, and T. Jungwirth, Rev. Mod. Phys. {\bf 87}, 1213 (2015).
  \bibitem{Silsbee1979} R. H. Silsbee, A. Janossy, and P. Monod, Phys. Rev. B {\bf 19}, 4382 (1979).
  \bibitem{Tserkovnyak2002} Y. Tserkovnyak, A. Brataas, and G. E. W. Bauer, Phys. Rev. Lett. {\bf 88}, 117601 (2002); Phys. Rev. B {\bf 66}, 224403 (2002).
  \bibitem{Uchida2008} K. Uchida, S. Takahashi, K. Harii, J. Ieda, W. Koshibae, K. Ando, S. Maekawa, and E. Saitoh, Nature {\bf 455},  778 (2008).
  \bibitem{Matsuo2013} M. Matsuo, J. Ieda, K. Harii, E. Saitoh, and S. Maekawa, 
Phys. Rev. B {\bf 87}, 180402(R) (2013).  
  \bibitem{Takahashi2016} R. Takahashi, M. Matsuo, M. Ono, K. Harii, H. Chudo, S. Okayasu, J. Ieda, S. Takahashi, S. Maekawa and E. Saitoh, Nat. Phys. {\bf 12}, 52 (2016).
  \bibitem{Kobayashi2017} D. Kobayashi, T. Yoshikawa, M. Matsuo, R. Iguchi, S. Maekawa, E. Saitoh, and Y. Nozaki, Phys. Rev. Lett. {\bf 119}, 077202 (2017).
  \bibitem{Funato2018} T. Funato and H. Kohno, J. Phys. Soc. Jpn. {\bf 87}, 073706 (2018).
  \bibitem{Kawada2021} T. Kawada, M. Kawaguchi, T. Funato, H. Kohno, and M. Hayashi, Sci. Adv. {\bf 7}, eabd9697 (2021).
  \bibitem{Rashba1959} E. I. Rashba and V. I. Sheka, Fiz. Tverd. Tela: Collection of Articles II, 162 (1959). 
 Also available as the supplementary material to: 
G. Bihlmayer, O. Rader, and R. Winkler, New. J. Phys. {\bf 17}, 050202 (2015). 
  \bibitem{Funato2021H} T. Funato and M. Matsuo, Phys. Rev. B {\bf 104}, L060412 (2021).
  \bibitem{Nagano2009} M. Nagano, A. Kodama, T. Shishidou, and T. Oguchi, J. Phys.: Condens. Matter. {\bf 21}, 064239 (2009).
  \bibitem{Yanase2011} Y. Yanase and H. Harima, \textit{Kotai Butsuri} (Solid State Physics) {\bf 46}, 283 (2011) [in Japanese].
  \bibitem{Slater1954} J. C. Slater and G. F. Koster, Phys. Rev. {\bf 94}, 1498 (1954). 
  \bibitem{Harrison1980} W. A. Harrison, \textit{Electronic Structure and Properties of Solids} (Freeman, San Francisco, 1980).
  \bibitem{Froyen1979} S. Froyen and W. A. Harrison, Phys. Rev. B {\bf 20}, 2420 (1979). 
  \bibitem{suppl} See Supplemental Material at (URL) for detailed calculations. 
  \bibitem{Park2013}  This feature was noted for $d$ bands in:   
J.-H. Park, C.-H. Kim, H.-W. Lee, and J.-H. Han, Phys. Rev. B {\bf 87}, 041301(R) (2013).
  \bibitem{com2} More microscopic treatment is possible by considering the long-range Coulomb interaction 
among electrons and background ions in the random phase approximation, as done in: 
 T. Funato and H. Kohno, Phys. Rev. B {\bf 102},  094426 (2020). 
  \bibitem{com1} Here, some errors in Ref.~\cite{Funato2021H} have been corrected, 
   and a generalization is made to allow effective-mass anisotropy. 
  \bibitem{com3} $\Xi_3$ and $2\Xi_4$ form a doublet (2D irreducible representation) of C$_{\infty v}$, 
whereas each forms a 1D representation of C$_{4v}$. 
  The same applies to pairs, $(\tau_3, \tau_1)$,  $(k_x^2-k_y^2, 2k_xk_y)$, and $(2\phi_3', \phi_4')$. 
  \bibitem{com4} At the same time, we assume 
$\omega \tau, q v_{\rm F} \tau \ll 1$ (diffusive regime). 
  \bibitem{Ishizaka2011}  K. Ishizaka, M. S. Bahramy, H. Murakawa, M. Sakano, T. Shimojima, T. Sonobe, 
K. Koizumi, S. Shin, H. Miyahara, A. Kimura, K. Miyamoto, T. Okuda, H. Namatame, M. Taniguchi, R. Arita, 
N. Nagaosa, K. Kobayashi, Y. Murakami, R. Kumai, Y. Kaneko, Y. Onose, and Y. Tokura, 
Nat. Mater. {\bf 10}, 521 (2011). 
  \bibitem{Rashba2003} E. I. Rashba, Phys. Rev. B {\bf 68}, 241315(R) (2003).
 \bibitem{com5} We note that the \lq\lq helicity current'' in Ref.~\cite{Funato2021H} corresponds 
to our quadupolar spin current. 


\end{thebibliography}
\end{document}